\documentclass[aps,preprint,showpacs,floatfix,pra]{revtex4}
\usepackage{bbding}
\usepackage{epsfig}
\usepackage{graphicx}
\usepackage{dcolumn}
\usepackage{bm}
\usepackage{ulem}
\usepackage{amssymb}
\usepackage{amsmath}   
 
\begin{document}
\title{ Current and entanglement in a Bose-Hubbard lattice}
\author{L. Morales-Molina, S. A. Reyes and M. Orszag}
\affiliation{ Facultad de F\'isica, Pontificia Universidad Cat\'olica de Chile, Casilla 306, Santiago 22, Chile}

\begin{abstract}
We study the generation of entanglement for interacting cold atoms in an optical lattice. The entanglement is generated by managing the interaction between two distinct atomic species. It is found that the current of one of the species can be used as a good indicator of entanglement generation. The thermalization process between the species is also shown to be closely related to the evolution of the current.
\end{abstract}

\maketitle

Entanglement is an important ingredient for the development of quantum information processing. Its study has been stimulated by its potential implementation in  quantum computing algorithms, making it ubiquitous in quantum systems, becoming particularly interesting in many-body systems \cite{Amico0}. Fast information processing is expected to involve many elements in the system, making particularly attractive the potential use of ultracold atoms. On the other hand, cold atom systems have proven to be ideal devices for ultra-high precision measurements \cite{Ober0}. Along these lines, several works involving entanglement in  ultracold atoms have been put forward \cite{Jaksch,Ober1,Ober2}.

In the present work we study a system where entanglement not only can be efficiently  generated, but also accurately estimated using current of particles. Here we consider that the parts of the system are two sets of different species of atoms moving on a lattice with tunable interaction. A sudden interaction quench between the two distinct atomic species is applied in order to generate entanglement.
In cold atom systems interaction can be easily manipulated, representing a great advantage with respect to semiconductor superlattices. Such a freedom to manage the interaction has been demonstrated experimentally \cite{FR} and has proven to be fundamental in phenomena such as coherent destruction of tunneling in a
many-body system \cite{gong}, control of atomic localization along a lattice \cite{morales}, stabilization of  
Bloch oscillations for a train of solitons in Bose-Einstein condensates (BEC) \cite{Gaul}, among others. On the other hand, increasing the interaction between atoms modifies the flux of particles along the lattice, giving rise in the limit of large interaction to a quantum phase transition from a superfluid state to a Mott insulator \cite{Amico}. Studies involving atoms moving along a ring shaped optical lattice have exploited such interaction combined with magnetic fields to generate a current of particles \cite{Currentflux}, as well as for the implementation of an AC  quantum motor \cite{denisov0}. 

We study a system where initially two sets of different species move independently along a ring with a finite current. As the interaction is turned on it not only modifies the currents, but also induces entanglement between 
both groups of atoms. We investigate the correlation between current and entanglement where the detection of the particle current variation provides a good measure of entanglement generation.
Interacting species of condensed atoms can be achieved either by using two kind of atoms, such as $^{87}$Rb and $^{40}$K \cite{Fesch} or from the same isotope with different internal states \cite{Ober2}.
The Hamiltonian for two interacting species of ultra-cold atoms reads 

\begin{equation} \label{Eq:Hamiltonian}
\hat{H}=\hat{H}_A\otimes \hat{\mathbf{1}}_{B}+\hat{H}_B \otimes \hat{\mathbf{1}}_{A}+ \hat{H}_{int},
\end{equation}
where the Hamiltonian for each specie $D=A,B$ is described by the Bose-Hubbard model  

\begin{equation}
\hat{H}_D=-C\sum_{j=1}^{L} (e^{i\phi/L} \hat{d}_{j}^{\dagger}\hat{d}_{j+1}+ e^{-i\phi/L} \hat{d}_{j+1}^{\dagger}\hat{d}_j)
+\frac{V}{2}\sum_{j=1}^{L}\hat{n}^D_{j}(\hat{n}^D_{j}-1),
\end{equation}
with $\hat{d}_{j}^{\dagger}$ ($\hat{d}=\hat{a},\hat{b}$) the creation and annihilation operator for the particles at site  $j$, $\hat{n}^D_{j}=d_j^{\dagger}d_{j}$ is the particle number operator and $L$ is the total number of sites. The first term describes the hopping or tunneling between adjacent sites in the lattice with a tunneling strength $C$. Here the hopping appears modified by a twist factor $e^{i\phi/L}$  which is generated via a magnetic field and produces a finite current along the ring \cite{Amico} . The above tigh-binding Hamiltonian is valid within the weak-coupling regime, that is, for optical lattices with a  depth $\gtrsim 5 E_{r}$, where $E_{r}=\hbar^2 k^2/2m$ is the single photon recoil energy  and $m$ is the atomic mass \cite{Ober}. Hereon we assume the same mass for both atomic species. The second term characterizes the on-site interaction with strength $V$ which, for simplicity, we consider equal to zero in what follows. 

The interspecies interaction Hamiltonian is given by 
\begin{equation}\label{Eq:HamSpecie}
H_{int}=U \sum_{j=1}^{L}  \hat{n}^A_{j} \otimes\hat{n}^B_{j},
\end{equation}
where $U$ denotes the interspecies interaction strength. Interspecies  interaction can be manipulated around Feschbach resonances \cite{Fesch} allowing great independence in tuning the interactions. 
Management of interactions provides a valuable tool for the control of the dynamics including generation of entanglement. 
Since we are interested in how the entanglement is generated between the species, we investigate the scenario with tunable interspecies interaction $U(t)$.
The full Hilbert space is spanned by the direct product of the single Fock states $|m_{A}\rangle\otimes |m_{B}\rangle$ with dimension $N_A\times N_B$. 
For three lattice sites the dimension of the Hilbert space for one  of the species is  $N_D=(n_D+1)(n_D+2)/2$, where $n_D$ is the corresponding number of atoms. 

Since we want to study the transport in a ring we use the periodic boundary condition $|L+1\rangle=|1\rangle$.
The current operator for the particles in the ring is $\hat{J}=\hat{J}_A\otimes \hat{\mathbf{1}}_{B}+ \hat{\mathbf{1}}_{A}\otimes \hat{J}_{B}$,
where $\hat{J}_{A,B}$ are the respective current operators for both species, defined as the sum of flux difference of particles between adjacent sites \cite{current}

\begin{equation}
\hat{J}_{D}=-\frac{iC}{\hbar L}\sum_{j=1}^{L}(e^{i\phi/L} \hat{d}_{j}^{\dagger}\hat{d}_{j+1}- e^{-i\phi/L} \hat{d}_{j+1}^{\dagger}\hat{d}_j).
\end{equation}

A good indicator of the degree of entanglement in a bi-partite pure state is provided by the Schmidt number ${\cal K}_0$ defined as the reciprocal of the purity of the reduced density matrix \cite{Miguel}
\begin{equation}\label{Eq:Schmidt0}
{\cal K}_0=\frac{1}{Tr_B(\rho_{B}^2)}.
\end{equation}
 For the sake of convenience, we define the normalized Schmidt number 
\begin{equation}\label{Eq:Schmidtnumber}
{\cal K}=({\cal K}_0-1)/(\Delta-1)\,\ \hbox{ where}\,\,\ 0\le {\cal K}\le 1
\end{equation}
and $\Delta=min(N_A,N_B)$ is the dimension of the system,  so the maximum degree of entanglement is found at ${\cal K}=1$.

 We consider an $L=3$ ring where the interspecies interaction is initially off and the cloud of particles  is prepared in the ground state of the optical lattice by using adiabatic loading techniques.
 Particles are then set in motion by tuning a magnetic field \cite{Currentflux}. Under these circumstances the Hamiltonian is reduced to the tight-binding description where one can  easily find the single particle eigenstates 
\begin{equation}
d^{\dagger}_{k}|0\rangle=\frac{1}{\sqrt{L}}\sum_{m=1}^{L}e^{imk} d^{\dagger}_{m}|0\rangle,
\end{equation}
where $|0\rangle$ is the vacuum state and $k_s=2\pi s/L$, $s=0,1,L-1$. Thus, the ground state of the non-interacting system with $n_A$ particles of type $A$ and $n_B$ particles of type $B$ becomes
\begin{equation}
|\psi_0 \rangle=\frac{1}{3^{(n_A+n_B)/2}} \frac{1}{\sqrt{n_{A}! n_{B}!}}\left(\sum_{m}e^{imk_A}a^{\dagger}_{m}\right)^{n_A}\left(\sum_{l}e^{ilk_{B}}b^{\dagger}_{l}\right)^{n_B}|0\rangle
\end{equation}
where $k_A$ ($k_B$) is the wavevector corresponding to the single particle ground state for $A$ ($B$) particles.
To get the two subsystems entangled, in analogy to experiments for the generation of entanglement using a nonlinear crystal, we consider the scenario where an attractive interaction strength is  suddenly turned on at $t=0$. Thus, we have a time-independent Hamiltonian before and after the interaction with constant interaction amplitude.  

\begin{figure}
 \begin{center}
\includegraphics[width=7.cm,height=6.cm]{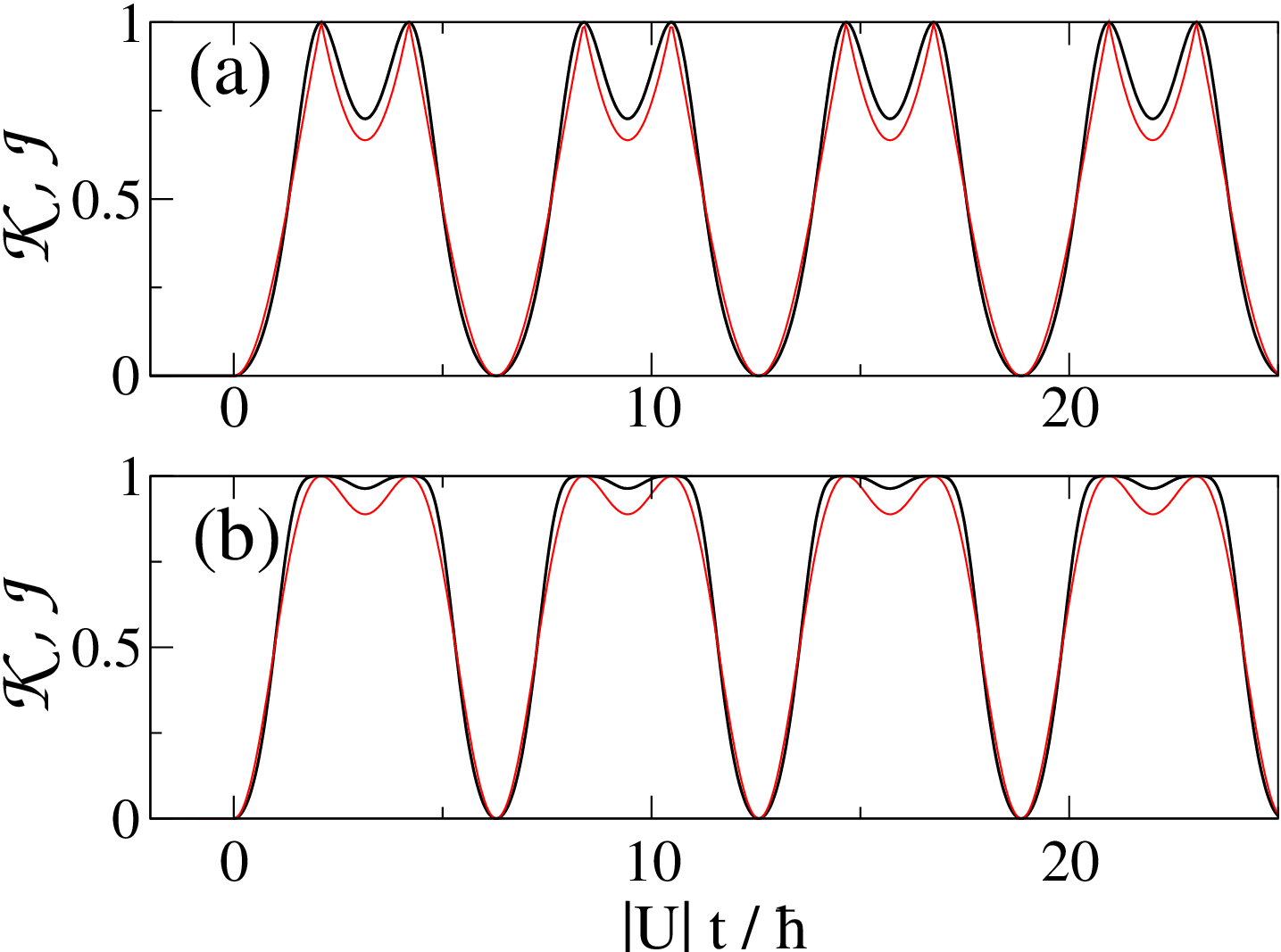}
\caption{(Color online) 
Evolution for attractive interaction $U=-10^4 C$. Black line: Schmidt number ${\cal K}$. Red line: Normalized current ${\cal J}$. In both panels the exact analytical expression is superimposed. (a) one particle in A and one particle in B; (b) two particles in A and one particle in B.}
\label{Fig:zeroCoupling}
\end{center}
\end{figure}

Let us consider first  the strong interaction limit $|C/U| \rightarrow 0$, such that at times $t>0$ the dynamics becomes determined by the interspecies interaction $H_{int}$. Thus, for a constant interaction, the wavefunction evolution is given by $e^{-i\frac{H_{int}}{\hbar}t}|\psi_{0}\rangle$. 
From now on, we set a small number of particles for the subsystem $B$, and an arbitrary number of particles for the subsystem $A$. For instance taking $n_B=1$ and $n_A=N$ results in the evolved wavefunction 
\begin{equation}\label{Eq:Wavefunction}
|\psi(t)\rangle=\frac{1}{3^{(N+1)/2}\sqrt{N!}} \sum_{l_1,...,l_2, m} e^{imk_B} e^{ik_A\sum_{j=1}^N l_j} e^{i\frac{Ut}{\hbar}(\sum_{j=1}^{N}\delta_{l_{j},m})} \prod_{j=1}^{N} a^{\dagger}_{l_j} b^{\dagger}_{m}|0\rangle
\end{equation}
Using Eq. \ref{Eq:Wavefunction} we find the density matrix  for the composite system $\rho_{AB}=|\psi(t)\rangle\langle\psi(t)|$ which after tracing over the system $A$ yields the reduced density matrix 
\begin{eqnarray}\label{Eq:traceB}
\rho_B(t) =Tr_A\{\rho_{AB}\}=\frac{1}{3^{N+1}}\sum_{n1+n2+n3=N, m,m'}
\left[\frac{N!}{n_1! n_2! n_3!}\right]^2 e^{ik_B(m-m')}\times
\nonumber \\ e^{i\frac{Ut}{\hbar}(\sum_{j=1}^{N}\delta_{l_{j},m}-\sum_{j=1}^N\delta_{l'_{j},m} )} b^{\dagger}_{m}|0\rangle \langle 0|b_{m'}.
\end{eqnarray}
This result can then be substituted into Eq.\ref{Eq:Schmidt0}  to estimate the entanglement. Exact calculations for the Schmidt number in the scenario of few particles can be found. In particular,  for $N=1$, we find
\begin{equation}\label{Eq:entanglement}
 {\cal{ K}}_{0}=\frac{27}{15+8\cos(Ut/\hbar)+4\cos(2Ut/\hbar)}
\end{equation}
 which maximizes at $2\pi/3+2s\pi$, and $4\pi/3+2s\pi$ with $s=0,1,2...$. The evolution for the normalized Schmidt number is shown in Fig. \ref{Fig:zeroCoupling}a. Interestingly,  a short time after the interaction is turned on the entanglement increases and reaches its maximal value. As shown in the figure, the entanglement remains high for a lapse of time before going down to zero again, repeating periodically with a time scale determined by $U$. 
If after some time the interaction is turned off, the entanglement of the system is preserved. 
Similar features are observed for $N=2$ in Fig. \ref{Fig:zeroCoupling}(b), where we also note an enhancement of the windows with  high entanglement.

Let us analyze now what happens with the current of the subsystem B at $t> 0$.  Considering again the evolved wavefunction $|\psi(t)\rangle$  for $N=1$ we find
\begin{equation}\label{Eq:analyticCurrent}
J_B(t) = \langle \hat{J}_B\rangle=\frac{2C}{9}\sin(k_{B}+\phi)\left(1+2\cos(Ut/\hbar)\right),
\end{equation}
which is a periodic function of the phase $\phi$.  For $t> 0$, the current becomes a time dependent function with a single frequency $U$.
Interestingly, $|J_{B}|$ goes down as the entanglement increases (see Eq.\ref{Eq:entanglement}). Moreover, the maximal entanglement points exactly match with the zeroes of Eq.\ref{Eq:analyticCurrent}.  In fact, it is convenient to introduce the normalized current
\begin{equation}
 {\cal{J}}(t) = 1 - \left|\frac{J_B(t)}{J_B(0)}\right|,
\end{equation}
which shows surprising similarity with the evolution of entanglement ${\cal K}(t)$ (see Fig. 1(a)). Such a result is interesting by itself, suggesting the use of current as a witness of the entanglement.  

\begin{figure}
 \begin{center}
\includegraphics[width=6.0cm,height=5.5cm]{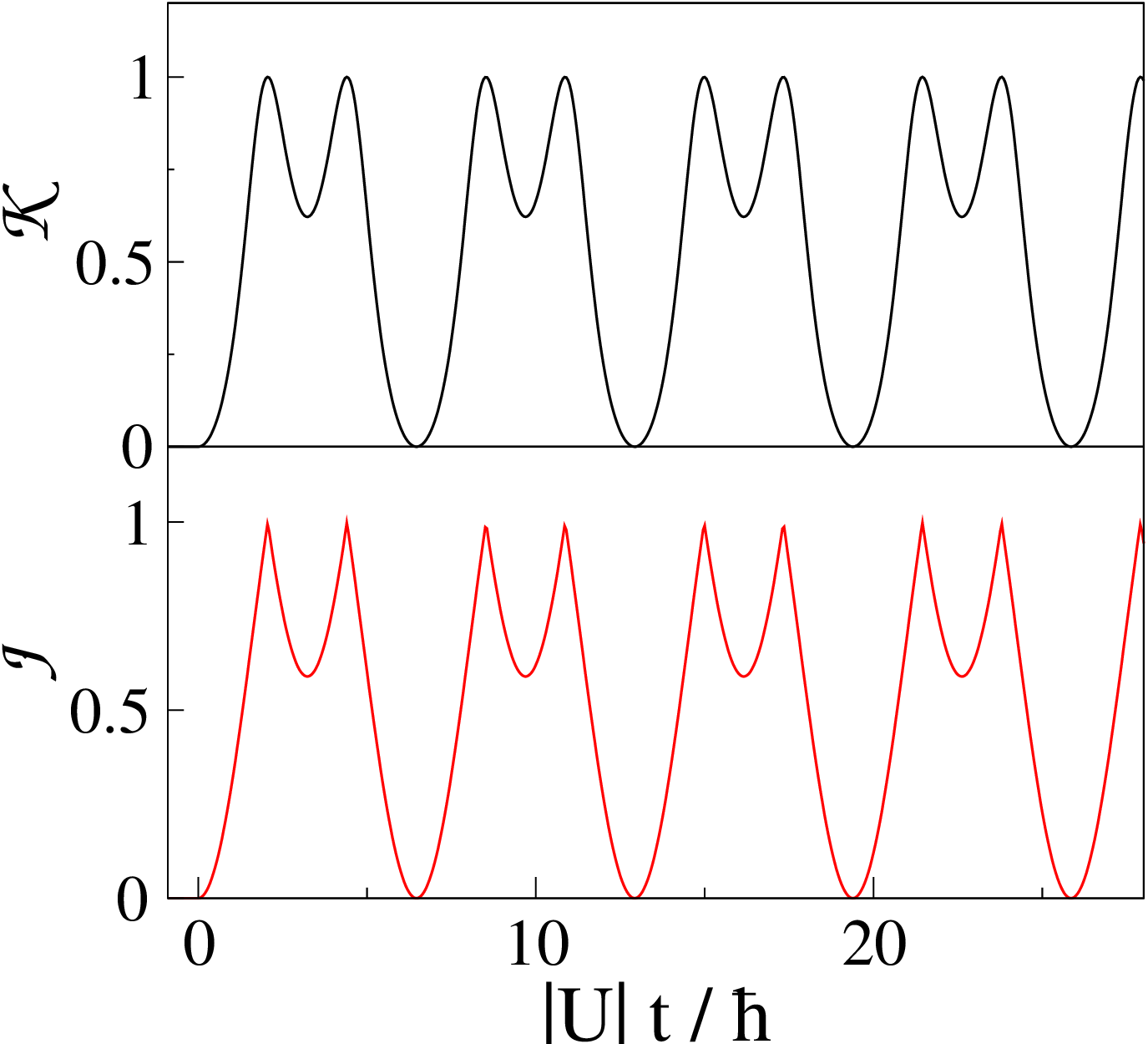}~~~
\includegraphics[width=6.0cm,height=5.5cm]{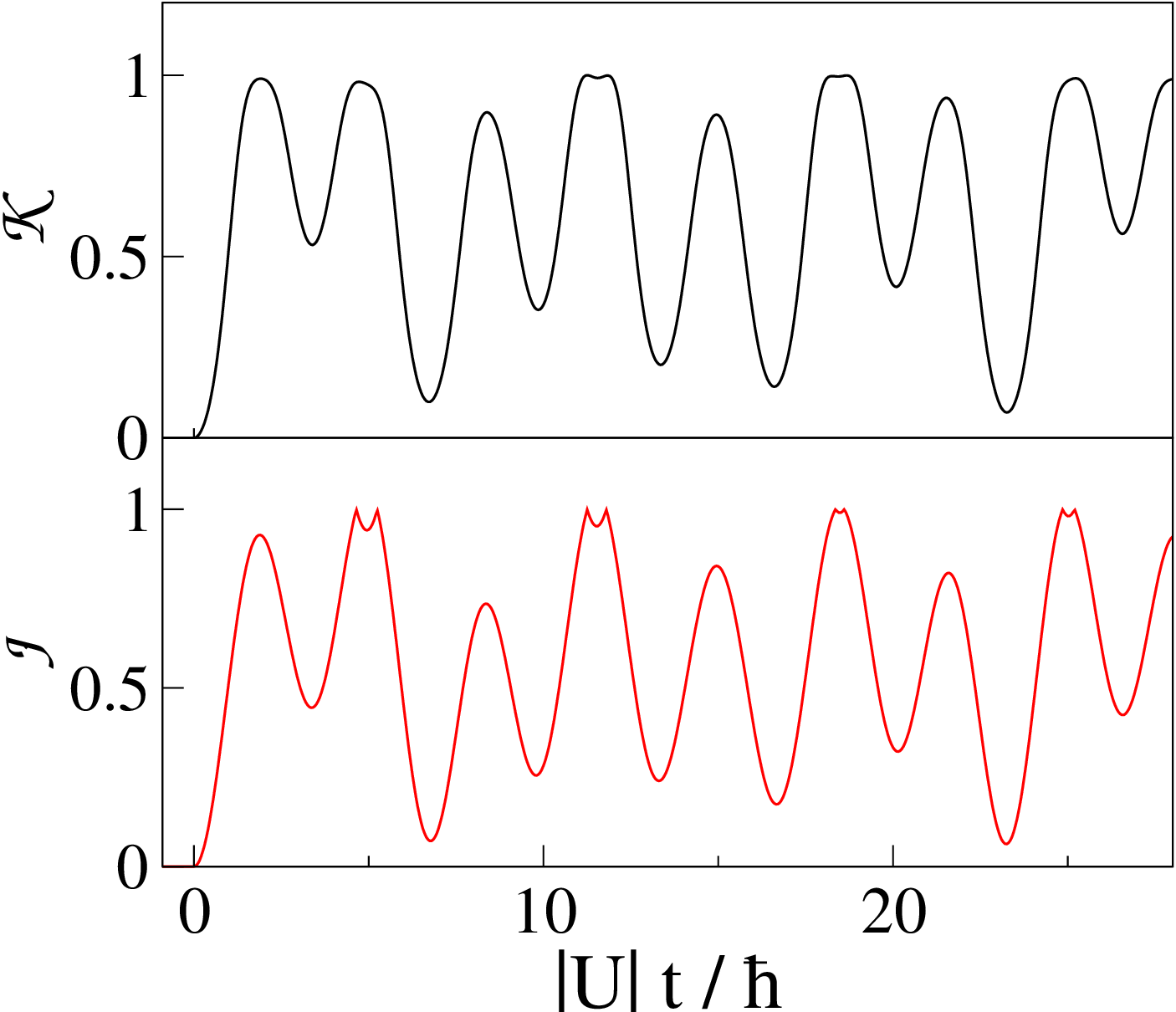}
\caption{Left and right panels: (Color online) Upper graph: Normalized Schmidt number ${\cal K}$ vs. time. Lower graph: ${\cal J}$ vs. time. The parameters are $U=-10 C$ and $\phi=\pi/10$. Left: one particle in A and one particle in B.
Right: two particles in A and one particle in B.}
\label{Fig:finiteCoupling}
\end{center}
\end{figure}

Contributions owing to the hopping elements in Eq. \ref{Eq:Hamiltonian} have been neglected so far. 
 To complement our previous analysis, the presence of a small contribution of the tigh-binding Hamiltonian is now considered within the strong interaction regime  $|C/U| \ll 1$. 
 As shown in Fig. \ref{Fig:finiteCoupling}(a), the entanglement evolution between one particle in each subsystem exhibits a similar behavior as for the zero hopping scenario, preserving the strong correlation between  ${\cal K}$ and ${\cal J}$. Indeed, we observe again that the current vanishes at the points of maximal entanglement and that the absolute value of the current decreases as entanglement increases. However, increasing the number of particles enhances the role of the hopping term generating fluctuations of the previous profile, as depicted in Fig.
\ref{Fig:finiteCoupling}(b).  Nevertheless, the normalized current of subsystem $B$ still shows signs of strong correlation with the entanglement.

\begin{figure}
 \begin{center}
\includegraphics[width=7.5cm,height=6.5cm]{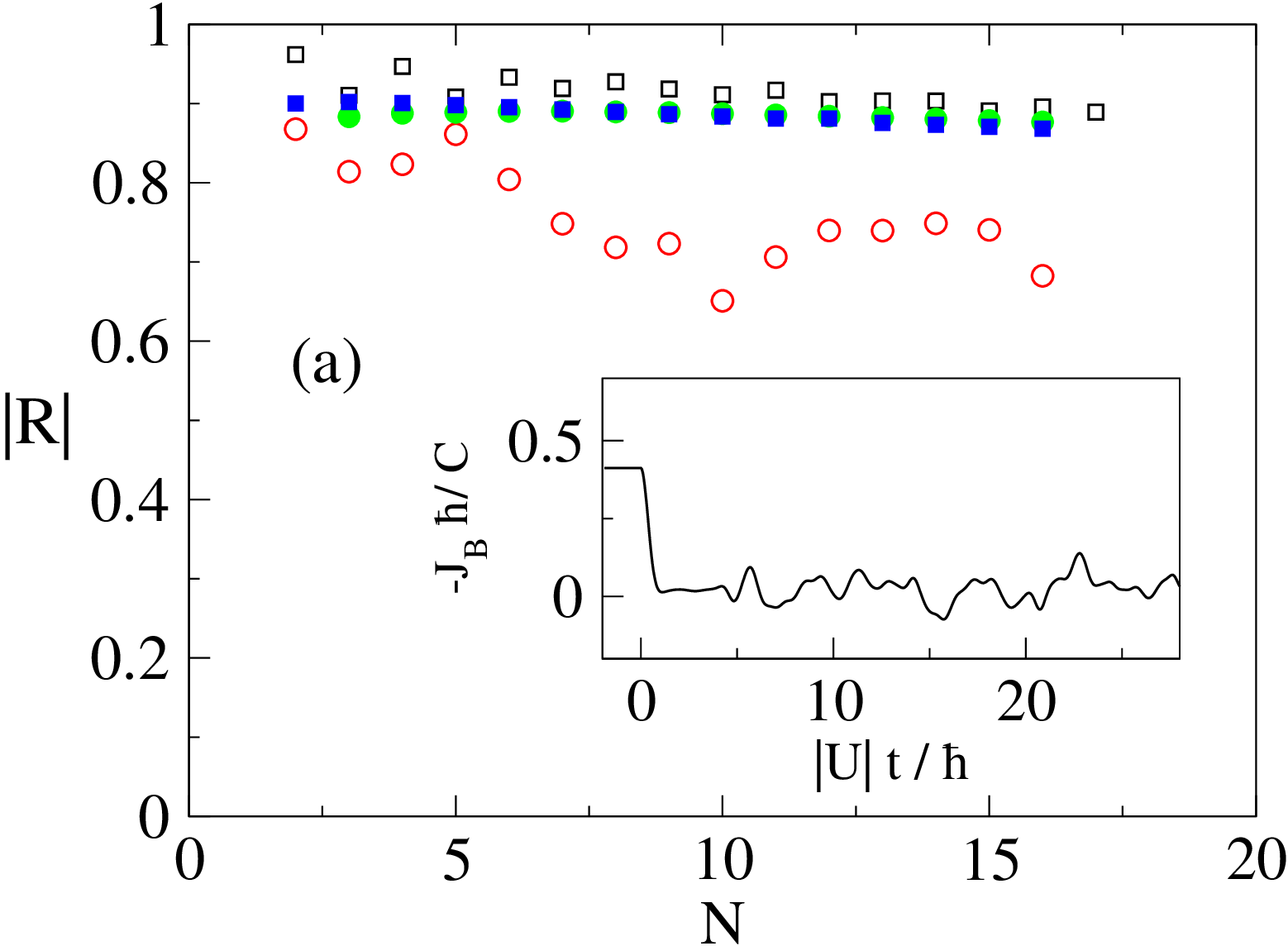}
\includegraphics[width=7.5cm,height=6.5cm]{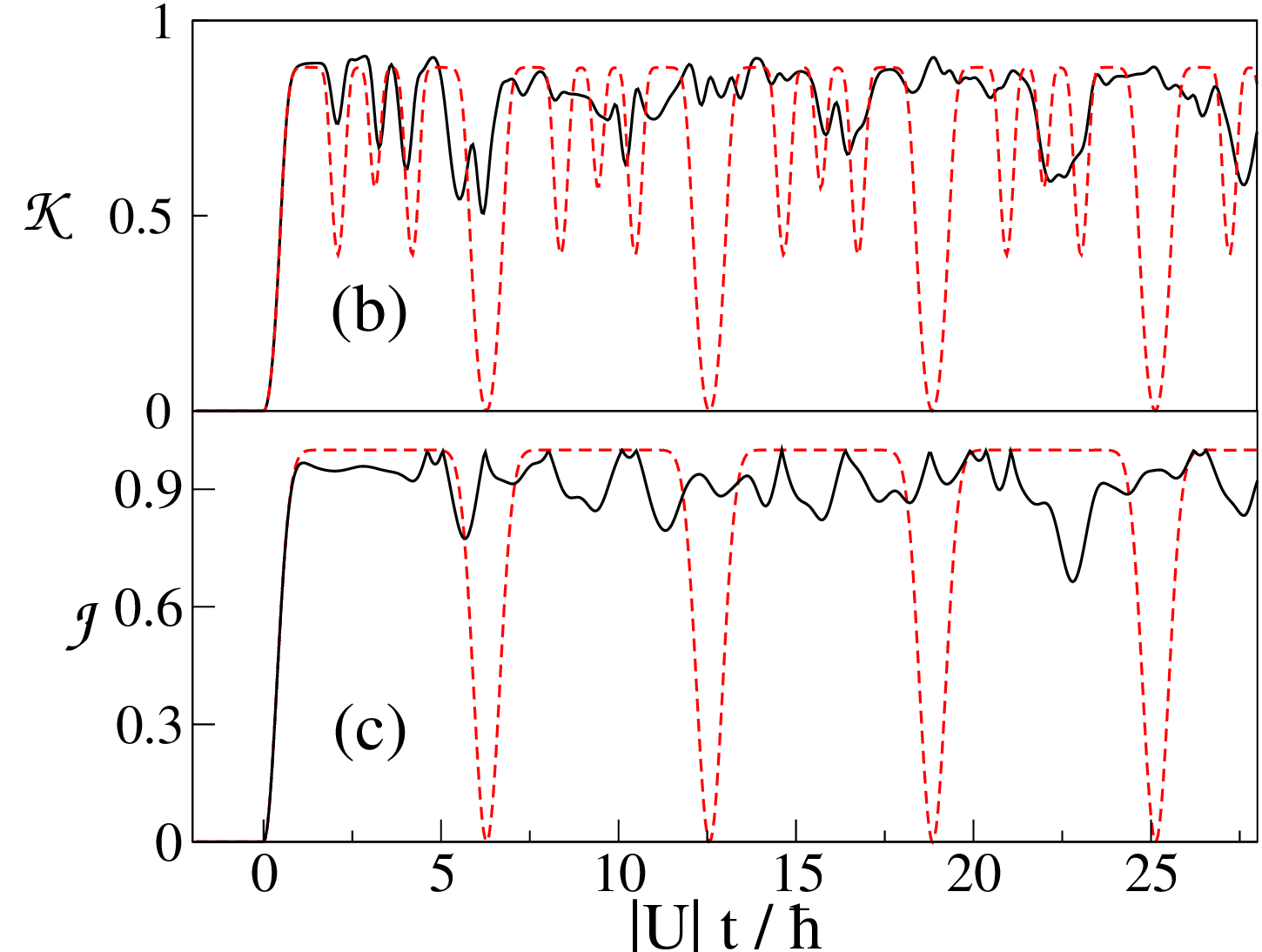}
\caption{(Color online) (a) Pearson coefficient vs. $N$. Squares and circles correspond to the scenario of one and two particles in $B$, respectively. Empty and filled symbols correspond to the $\tau_{s}$ and $\tau_{l}$ time scales, respectively. Inset shows the evolution of the current of two particles in $B$ in units of $C/\hbar$ for $N=11$.
(b) Normalized Schmidt number ${\cal K}$ vs. time. (c) ${\cal J}$ vs.time. In (b) and (c)
dashed-line: $C/|U|= 10^{-4}$, solid line: $C/|U|= 1/10$ .The other parameters are the same as in Fig.\ref{Fig:finiteCoupling}.
}
\label{Fig:Coupling2}
\end{center}
\end{figure}

 We are interested in estimating the aforementioned correlation within  a short timescale $\tau_s=2\tau_U$ and a much larger one $\tau_l=20\tau_U$, where $\tau_{U}\equiv \hbar/|U|$ is the characteristic time of the system.
To quantify the degree of correlation between the normalized current and the normalized entanglement, we consider the Pearson product-momentum coefficient number  
\begin{equation}
 R=\frac{\textstyle{E[({\cal{K}}-\mu_{\cal{K}})({\cal J}-\mu_{{\cal J}})]}}{\textstyle{\sigma_{\cal{K}}\sigma_{{\cal J}}}},
\end{equation}
where $\mu$'s and $\sigma$'s are the mean values and standard deviations respectively. The absolute value of the coefficient $R$ is close to unity when the variables involved are highly correlated and approaches zero when there is no correlation.
Fig.\ref{Fig:Coupling2}(a) exhibits the results obtained from numerical simulations for two distinct sets of atoms number one and two particles in subsystem $B$ and a finite hopping $C=|U|/10$. For $n_B=1$ the correlation between ${\cal{K}}$ and ${\cal{J}}$ remains high even when the number of particles in subsystem $A$ goes up to $N=17$ for both time scales considered.  Furthermore, for $n_B=2$ a high degree of correlation appears to be almost independent of $N$ for the short time scale $\tau_{s}$,  whereas for a much larger time $\tau_{l}$ the correlation decreases when the number of particles increases. 

Remarkably, for the scenario with large number of particles in $A$ the entanglement does not decrease significantly after its initial growth as shown in Fig. \ref{Fig:Coupling2}(b) for $N=11$. Instead, the Schmidt number remains high, displaying only small fluctuations around its average value.
This behavior arises because the finite hopping between adjacent sites lifts the degeneracies of $H_{int}$ such that the small subsystem $B$ can now equilibrate with the much larger system $A$. In other words, particles of type $A$ play the role of a thermal bath for system $B$.

Lastly, we observe that as the number of particles in $A$ grows, the current of the subsystem with few particles tends to evolve towards an equilibrium value very close to zero, corresponding to ${\cal J}\approx 1$ (see Fig. \ref{Fig:Coupling2}(c)). 
To understand this let us analyze the current of particles in the subsystem $B$ 
\begin{equation}
 J_B = \frac{1}{Tr(\hat{\rho_{B}})}\sum_{m,m'}\langle m'|\hat{\rho}_{B}|m \rangle \langle m |\hat{J}_{B}|m'\rangle
\end{equation}
where $|m\rangle=b_m^{\dagger}|0\rangle$.
Now, since the only non-zero matrix elements of the current are $\langle m |J_{B}|m\pm 1\rangle$ then only off-diagonal elements of $\hat{\rho}_{B}$ would contribute to the current. Nevertheless, it is known that in the limit $n_{A}\gg n_{B}$ entanglement between particles in $B$ and $A$ results in a thermalization process \cite{Thermal}, where the non-diagonal elements of the reduced density matrix are reduced to zero \cite{denisov}. Thus, the observation that $J_B$ equilibrates to zero is a direct consequence of the nature of the relaxation dynamics. Remarkably, the current could be used not only as an indicator of entanglement generation but also of how close the system is to thermal equilibrium.

To conclude, we have studied the physical detection of entanglement for two sets of different atomic species moving in a ring-shaped optical lattice. The setup allows high entanglement generation and also permits the estimation of the degree of entanglement by measuring the flux of particles. We have found that the generation of entanglement is highly correlated to the modification of the current in one of the subsystems. This current could then be used as a ``witness'' of entanglement generation in future experiments. Furthermore, it was shown that the same current can be used as a tool for the detection of thermalization between species of atoms.

LMM and SR acknowledge financial support from FONDECYT
project no 1110671. SR is also supported by FONDECYT
project no 11110537 and MO  is supported by FONDECYT
project no 1100039.

\end{document}